\documentclass[11pt,a4paper]{article}
\usepackage{jcappub,natbib,epsfig,graphicx}
\usepackage{braket}
\title{CMB statistical anisotropy from noncommutative gravitational waves
}
\author[a,b]{Maresuke Shiraishi,}
\author[c]{David F. Mota,}
\author[a,b]{Angelo Ricciardone}
\author[b]{and Frederico Arroja}
\affiliation[a]{Dipartimento di Fisica e Astronomia ``G. Galilei'', Universit\`a degli Studi di Padova, \\ 
via Marzolo 8, I-35131, Padova, Italy}
\affiliation[b]{INFN, Sezione di Padova, \\ 
via Marzolo 8, I-35131, Padova, Italy}
\affiliation[c]{Institute of Theoretical Astrophysics, University of Oslo, \\ 
P.O. Box 1029 Blindern, N-0315 Oslo, Norway}

\abstract{%
Primordial statistical anisotropy is a key indicator to investigate early Universe models and has been probed by the cosmic microwave background (CMB) anisotropies. In this paper, we examine tensor-mode CMB fluctuations generated from anisotropic gravitational waves, parametrised by $P_h({\bf k}) = P_h^{(0)}(k) [ 1 + \sum_{LM} f_L(k) g_{LM} Y_{LM} (\hat{\bf k}) ]$, where $P_h^{(0)}(k)$ is the usual scale-invariant power spectrum. Such anisotropic tensor fluctuations may arise from an inflationary model with noncommutativity of fields. It is verified that in this model, an isotropic component and a quadrupole asymmetry with $f_0(k) = f_2(k) \propto k^{-2}$ are created and hence highly red-tilted off-diagonal components arise in the CMB power spectra, namely $\ell_2 = \ell_1 \pm 2$ in $TT$, $TE$, $EE$ and $BB$, and $\ell_2 = \ell_1 \pm 1$ in $TB$ and $EB$. We find that B-mode polarisation is more sensitive to such signals than temperature and E-mode polarisation due to the smallness of large-scale cosmic variance and we can potentially measure $g_{00} = 30$ and $g_{2M} = 58$ at 68\% CL in a cosmic-variance-limited experiment. Such a level of signal may be measured in a PRISM like experiment, while the instrumental noise contaminates it in the {\it Planck} experiment. These results imply that it is impossible to measure the noncommutative parameter if it is small enough for the perturbative treatment to be valid. Our formalism and methodology for dealing with the CMB tensor statistical anisotropy are general and straightforwardly applicable to other early Universe models.
}
\emailAdd{maresuke.shiraishi@pd.infn.it}
\emailAdd{d.f.mota@astro.uio.no}
\emailAdd{angelo.ricciardone@pd.infn.it}
\emailAdd{arroja@pd.infn.it}

\begin{document}

\maketitle
\flushbottom

\section{Introduction}
 
Noncommutativity in quantum fluctuations may be a key property of fundamental theories at ultra high energy scales, such as string theory and Lorentz-violating theories (see e.g., refs.~\cite{Seiberg:1999vs, Seiberg:2000gc, mota1,mota2,Tsujikawa:2003gh, Ferrari:2006gs}). These theories have been well-studied for the purpose of solving several open issues in cosmology and high energy physics such as magnetogenesis (e.g., refs.~\cite{Mazumdar:2000jc, li:2007,Gamboa:2005bf, Kouretsis:2013cga,bertolami}). 

Quantum fluctuations affected by noncommutativity in the very early Universe can imprint their characteristic signatures in cosmological observables. A space-time noncommutativity can induce primordial curvature perturbations, which breaks rotational and parity invariance \cite{Alexander:2001dr, Lizzi:2002ib, Koh:2007rx, Akofor:2007fv, Kobakhidze:2008cq, Koivisto:2010fk, Nautiyal:2013bwa}. These act as seeds of the cosmic microwave background (CMB) anisotropies and generate dipole or quadrupole anisotropy in the CMB power spectrum \cite{Lizzi:2002ib, Akofor:2007fv, Akofor:2008gv, Karwan:2009ic, Koivisto:2010fk}. Nowadays, primordial noncommutativity has attracted attention from both purely theoretical and phenomenological sides. 

In addition to the curvature perturbation, the noncommutativity also affects the tensor-mode sector, i.e, gravitational waves. In ref.~\cite{Cai:2007xr}, the impacts of tensor-mode noncommutativity on the present gravitational wave background (GWB) have been investigated. There, the authors have taken into account not the space-time noncommutativity but the noncommutativity of fields by the so-called noncommutative field approach. This is also motivated by the Lorentz-violating gravity \cite{Carmona:2002iv, Falomir:2005it, Ferrari:2006gs}. The resulting GWB originates from the inflationary tensor-mode perturbations, while the superhorizon modes of such seed perturbations should also create CMB tensor-mode fluctuations at late times. The present paper examines such CMB signatures for the first time. We find that, in the same manner as the scalar-mode case, resultant CMB power spectra violate rotational invariance since the noncommutativity of fields induces a preferred direction.

Probing statistical anisotropy in the CMB fluctuations is one of the most interesting and attractive topics in cosmology (e.g., refs.~\cite{Groeneboom:2009qf,axel, Groeneboom:2010fn, Ramazanov:2013wea}). Current CMB temperature data in the {\it Planck} experiment indicates nonzero dipole anisotropy \cite{Ade:2013nlj} and also gives the most accurate bound on the quadrupole anisotropy \cite{Kim:2013gka}. These results have been obtained from the CMB scalar-mode power spectra. On the other hand, meaningful signals can also arise in the tensor-mode sector as shown in the present noncommutative case. In this sense, probing the statistical anisotropy from the tensor-mode power spectra should be a beneficial study. In the present paper, we show how the tensor-mode statistical anisotropy can be measured in current and forthcoming experiments such as ${\it Planck}$ \cite{Ade:2013ktc} or PRISM \cite{Andre:2013afa, Andre:2013nfa}. 

To analyse the CMB signatures, at first, we derive the power spectrum of the inflationary gravitational wave by following ref.~\cite{Cai:2007xr}. Then, it is obvious that the noncommutativity produces the quadrupole anisotropy. We express its magnitude through the spherical harmonic coefficients, $g_{LM}$, given by   
\begin{eqnarray}
P_h({\bf k}) 
= P_h^{(0)}(k) \left[ 1 + 
 \sum_{L = 0}^\infty \sum_{M = -L}^{L} f_L(k)
g_{LM} Y_{LM} (\hat{\bf k}) \right] ~, \label{eq:def_gLM}
\end{eqnarray}
where $P_h^{(0)}(k)$ is the usual scale-invariant isotropic power spectrum. This parametrisation is similar to the scalar-mode case \cite{Pullen:2007tu}. In the present noncommutative case, $g_{LM}$ vanishes except in $L = 0$ and 2 due to parity conservation. Interestingly, $f_0(k) = f_2(k) \propto k^{-2}$ is found, and hence we observe a highly red-tilted power spectrum of the CMB fluctuations. Applying the quadratic maximum likelihood (QML) estimator utilised in weak lensing analysis \cite{Hanson:2009gu, Hanson:2010gu}, we compute the expected uncertainty on $g_{LM}$. Then, we perform joint analyses of the auto- and cross-spectra between temperature, E-mode and B-mode polarisations. As a consequence, we find that B-mode signals are more informative and can potentially measure $g_{00} = 30$ and $g_{2M} = 58$, where $M = 0, \pm 1, \pm 2$, at 68\% CL. 

The plan of this paper is as follows. In the next section, we derive the power spectrum of the inflationary gravitational waves. In section~\ref{sec:CMB}, we compute the resulting CMB power spectra and evaluate the expected uncertainty on $g_{LM}$ by using the QML estimator. The final section is devoted to the conclusion. 

\section{Noncommutative gravitational waves}

In this section, we estimate the power spectrum of inflationary gravitational waves under the presence of noncommutativity of fields. The model we discuss here has been firstly analysed by ref.~\cite{Cai:2007xr} for the purpose of the GWB measurement. Here, we shall compute the superhorizon-scale power spectrum of primordial gravitational waves, which creates tensor-mode power spectra of the CMB fluctuations in the late-time Universe.

\subsection{Noncommutative system}

Let us consider gravitational waves, $\bar{h}_{ij}$, on the FLRW background, namely $ds^2 = a^2(\tau) [-d\tau^2 + (\delta_{ij} + \bar{h}_{ij})dx^i dx^j]$, where $a(\tau)$ is the scale factor as a function of conformal time $\tau$ and $\partial_i \bar{h}_{ij} = \bar{h}_{ii} = 0$. We start from the standard quadratic action in the tensor-mode sector:  
\begin{eqnarray}
S &=& \frac{M_{\rm pl}^2}{8} \int d\tau d^3 x a^2 
\left[ \dot{\bar{h}}_{ij}^2 - (\partial_l \bar{h}_{ij})^2  \right] \nonumber \\ 
&=& \frac{1}{4} \int d\tau d^3 x 
\left[ \dot{h}_{ij}^2 + \frac{\ddot{a}}{a} h_{ij}^2 - (\partial_l h_{ij})^2  \right] ~,
 \label{eq:action}
\end{eqnarray}
where $~\dot{}~ \equiv \partial_\tau$ denotes a derivative with respect to conformal time and $M_{\rm pl} = 1/\sqrt{8\pi G}$ is the reduced Planck mass. To derive the second equality, we have introduced canonical normalisation $h_{ij} \equiv a M_{\rm pl} \bar{h}_{ij} / \sqrt{2}$. Then, we impose the noncommutativity of fields as\footnote{Here, we use the helicity-state representations, instead of usual Fourier-space ones, e.g., $\left[ h_{ij}({\bf k}, \tau), p_{kl}({\bf k'}, \tau) \right]$, seen in the original paper~\cite{Cai:2007xr}, because the expressions are simpler and the transverse and traceless conditions are automatically satisfied.}
\begin{eqnarray}
\begin{split}
\left[ h_{\bf k}^{(\lambda)}(\tau), h_{\bf k'}^{(\lambda')}(\tau) \right] &= 0 ~, \\
\left[ h_{\bf k}^{(\lambda)}(\tau), p_{\bf k'}^{(\lambda')}(\tau) \right] &= 
\frac{i}{2} (2\pi)^3 \delta^{(3)}({\bf k} + {\bf k'})
 \delta_{\lambda, \lambda'} ~, \\
\left[ p_{\bf k}^{(\lambda)} (\tau), p_{\bf k'}^{(\lambda')}(\tau) \right] &=
-\lambda {\boldsymbol \alpha} \cdot \hat{\bf k} (2\pi)^3 \delta^{(3)}({\bf k} + {\bf k'}) \delta_{\lambda, \lambda'}
 ~, 
\label{eq:nc_rel}
\end{split}
\end{eqnarray}
where $h_{\bf k}^{(\lambda)}$ and $p_{\bf k}^{(\lambda)} \equiv  \frac{1}{2} \dot{h}_{\bf k}^{(\lambda)}$ denote the $\lambda = \pm 2$ helicity-space expressions of the gravitational wave $h_{ij}$ and its conjugate momentum $p_{ij} = \delta {\cal L} / \delta \dot{h}_{ij} =  \frac{1}{2} \dot{h}_{ij}$ with ${\cal L}$ being the Lagrangian density of the corresponding action, and $h_{ij}$ is given by
\begin{eqnarray}
h_{ij}({\bf x}, \tau) 
&=& \int\frac{d^{3}{\bf k}}{(2\pi)^{3}}
\sum_{\lambda = \pm 2} h_{\bf k}^{(\lambda)}(\tau) 
e^{(\lambda)}_{ij}(\hat{\bf k}) e^{i{\bf k}\cdot {\bf x}} 
~.
\end{eqnarray}
The polarisation tensor $e_{ij}^{(\lambda)} (\hat{\bf k})$ satisfies $e_{ii}^{(\lambda)}(\hat{\bf k}) = \hat{k}_i e_{ij}^{(\lambda)}(\hat{\bf k}) = 0$ (transverse and traceless condition), $e_{ij}^{(\lambda)}(\hat{\bf k}) e_{ij}^{(\lambda')}(\hat{\bf k}) = 2 \delta_{\lambda, -\lambda'}$ (normalisation) and $e_{ij}^{(\lambda) *}(\hat{\bf k}) = e_{ij}^{(-\lambda)}(\hat{\bf k}) = e_{ij}^{(\lambda)}(- \hat{\bf k})$ \cite{Shiraishi:2010kd}. The so-called noncommutative parameter ${\boldsymbol \alpha}$ expresses the size and the direction of the noncommutativity of the conjugate momenta. Notice that the usual commutative relations are restored by ${\boldsymbol \alpha} = 0$. Physically, the magnitude of ${\boldsymbol \alpha}$ determines the comoving energy scale where the noncommutativity plays a significant role in graviton propagation. In the following discussions, for simplicity, let us analyze the phenomenological signatures of the noncommutativity in a special case: ${\boldsymbol \alpha} = {\rm const}$, although it may be possible to consider running ${\boldsymbol \alpha}$. 

\subsection{Quantization}

The present case imposes the noncommutativity between conjugate momenta \eqref{eq:nc_rel}. This noncommutative effect may be interpreted as an additional term breaking the Lorentz invariance in an effective action \cite{Cai:2007xr}:
\begin{eqnarray}
S^{\rm new} 
\equiv \frac{1}{4} \int d\tau d^3 x 
\left[ \dot{h}_{ij}^2 + \frac{\ddot{a}}{a} h_{ij}^2 - (\partial_l h_{ij})^2 
- 8 \alpha_m \eta_{jkm} h_{kl} \dot{h}_{lj} \right] ~,
 \label{eq:action_new}
\end{eqnarray}
with $\eta_{ijk}$ being the three dimensional antisymmetric tensor normalised as $\eta_{123} = 1$, and the usual commutation relations are restored, reading 
\begin{eqnarray}
\begin{split}
\left[ h_{\bf k}^{(\lambda)}(\tau), \pi_{\bf k'}^{(\lambda')} (\tau) \right] 
&= \frac{i}{2}  (2\pi)^3 
\delta^{(3)}({\bf k} + {\bf k'}) \delta_{\lambda, \lambda'}
~, \\ 
\left[ \pi_{\bf k}^{(\lambda)}(\tau), \pi_{\bf k'}^{(\lambda')} (\tau) \right] &= 0~.
\end{split} 
\label{eq:c_rel}
\end{eqnarray}
Here, the helicity-state expression of the new conjugate momentum $\pi_{ij} \equiv \delta {\cal L}^{\rm new} / \delta \dot{h}_{ij}$ is given by $\pi_{\bf k}^{(\lambda)} = \frac{1}{2} e_{ij}^{(-\lambda)}(\hat{\bf k}) \pi_{ij}({\bf k}) = p_{\bf k}^{(\lambda)} - i \lambda {\boldsymbol \alpha} \cdot \hat{\bf k} h_{\bf k}^{(\lambda)}$. This process is the so-called noncommutative field approach and, owing to eq.~\eqref{eq:c_rel}, we can perform the normal quantization process as described below. 

Quantized gravitational waves can be expressed as 
\begin{eqnarray}
h_{\bf k}^{(\lambda)}(\tau) 
= v_{\bf k}^{(\lambda)}(\tau) \beta({\bf k}, \lambda) 
+ v_{- {\bf k}}^{(\lambda) *}(\tau) \beta^\dagger(-{\bf k}, \lambda)~, \label{eq:GW_fourier}
\end{eqnarray}
where $~\hat{}~$ denotes a unit vector and the creation $\beta^\dagger$ and annihilation $\beta$ operators obey $\beta({\bf k}, \lambda)\Ket{0} = 0$ and $\left[ \beta({\bf k}, \lambda), \beta^\dagger({\bf k'}, \lambda') \right] = (2\pi)^3 \delta_{\lambda, \lambda'} \delta^{(3)}({\bf k} - {\bf k'})$, with $\lambda, \lambda' = \pm 2$. Using these relations, one can understand that the above commutation relations \eqref{eq:nc_rel} and \eqref{eq:c_rel} equate to the conditions on the mode function $v_{\bf k}^{(\lambda)}$: 
\begin{eqnarray}
\begin{split}
|v_{\bf k}^{(\lambda)}|^2  &= |v_{- \bf k}^{(\lambda)}|^2 ~, \\ 
v_{\bf k}^{(\lambda)} \dot{v}_{\bf k}^{(\lambda) *} 
- v_{-\bf k}^{(\lambda) *} \dot{v}_{-\bf k}^{(\lambda)} &= i ~, \\ 
|\dot{v}_{\bf k}^{(\lambda)}|^2 - |\dot{v}_{- \bf k}^{(\lambda)}|^2 
 &= - 4 \lambda {\boldsymbol \alpha} \cdot \hat{\bf k} ~.
\end{split} \label{eq:ini_commu}
\end{eqnarray}
These will be used to determine the normalisation of the mode functions.

\subsection{Tensor power spectrum}

The variation of the action (\ref{eq:action_new}) with respect to $h_{\bf k}^{(\lambda)}$ leads to the equation of motion:
\begin{eqnarray} 
\ddot{v}^{(\lambda)}_{\bf k}  
- 4 i\lambda {\boldsymbol \alpha} \cdot \hat{\bf k} 
 \dot{v}^{(\lambda)}_{\bf k} 
+ \left( k^{2} - \frac{\ddot{a}}{a} \right) v^{(\lambda)}_{\bf k} 
 = 0 ~, \label{eq:EOM}
\end{eqnarray}
where we have used an useful relation: $\eta_{ijk} e_{il}^{(\lambda)}(\hat{\bf k}) e_{jl}^{(\lambda')}(\hat{\bf k}) = -i \lambda \hat{k}_k \delta_{\lambda, -\lambda'}$ \cite{Shiraishi:2010kd}. Note that this form is valid in terms of any direction of ${\boldsymbol \alpha}$ and coincides with the result in ref.~\cite{Cai:2007xr} under the condition where ${\boldsymbol \alpha}$ is parallel to the $z$ axis. 

Let us work in standard slow-roll inflation, namely 
\begin{eqnarray}
\frac{\ddot{a}}{a} \simeq \frac{ \nu^{2}-\frac{1}{4}}{\tau^{2}} ~,
\end{eqnarray} 
where $\nu = \frac{3}{2} + \frac{\epsilon}{1-\epsilon}$ with $\epsilon$ being the slow-roll parameter. Then, by imposing eq.~\eqref{eq:ini_commu} as initial conditions on subhorizon scales, we can easily find a solution of eq.~\eqref{eq:EOM} as 
\begin{eqnarray}
v^{(\lambda)}_{\bf k} (\tau) = 
 \frac{\sqrt{\pi} }{2} e^{i(\frac{\nu \pi}{2} + \frac{\pi}{4})} 
\sqrt{-\tau } e^{2 i \lambda {\boldsymbol \alpha} \cdot \hat{\bf k} \tau } 
 H_\nu^{(1)}(-l\tau) ~,
\end{eqnarray}
where $H_\nu^{(1)}(x)$ is the Hankel function of the first kind and $l \equiv \sqrt{k^2+ 16 ({\boldsymbol \alpha} \cdot \hat{\bf k} )^2 }$. One can see from this form that the presence of the noncommutativity changes the phase and the dispersion relation of the mode function. The power spectrum of the original gravitational waves $\bar{h}_{ij}$ is straightforwardly given by
\begin{eqnarray}
\Braket{\prod_{i=1}^2 \bar{h}^{(\lambda_i)}_{{\bf k}_i}(\tau)}
&=& (2\pi)^3 \frac{P_h({{\bf k}_1},\tau)}{2} \delta_{\lambda_1, \lambda_2} 
\delta^{(3)}({{\bf k}_1} + {{\bf k}_2}) ~, \\
P_h({\bf k}, \tau) &=& \frac{4}{a^2 M_{\rm pl}^2} |v_{\bf k}^{(\lambda)}(\tau)|^2 ~.
\end{eqnarray}
Assuming de-Sitter like space-time ($\epsilon \approx 0$ or $\nu \approx \frac{3}{2}$), the superhorizon power spectrum which acts as a source of the CMB power spectra becomes
\begin{eqnarray}
P_h({\bf k}) 
\approx \frac{2 H^2}{M_{\rm pl}^2} 
 l^{-3} 
\approx \frac{2 H^2}{M_{\rm pl}^2} k^{-3} 
  \left[ 1 - 24 (\hat{\boldsymbol \alpha} \cdot \hat{\bf k})^2 \left(\frac{\alpha}{k}\right)^{2} \right] 
 ~, \label{eq:GW_power_n0}
\end{eqnarray}
where $H$ is the Hubble parameter. In the final approximation, we have treated the anisotropic part perturbatively. Together with the positiveness of the power spectrum, this restricts the values of the noncommutative parameter one can analyze to $|{\boldsymbol \alpha} \cdot \hat{\bf k}| \ll k / 5 $. The observational limit on the anisotropy of the scalar fluctuations suggests that this is a reasonable assumption also for the tensor mode \cite{Kim:2013gka, Ramazanov:2013wea}.

Interestingly, this form involves the quadrupole anisotropy depending on $k^{-5}$. It is a consequence of the quadratic dependence of $l$ on ${\boldsymbol \alpha} \cdot \hat{\bf k}$. Accordingly, the resulting CMB power spectra break the statistical isotropy and off-diagonal signals, namely $\ell_1 \neq \ell_2$, will arise. Note that parity invariance is kept since spin-$\lambda$ dependence cancels out.

\section{Signatures in the CMB anisotropies} \label{sec:CMB}

In this section, we shall discuss signatures of the anisotropic gravitational waves in the CMB temperature and polarisation power spectra, and we investigate the expected uncertainty on the magnitude of anisotropy from observations using the quadratic maximum-likelihood (QML) estimator \cite{Hanson:2009gu, Hanson:2010gu}. 

For convenience of analysis, let us express the anisotropic power spectrum of gravitational waves by the spherical harmonic expansion \eqref{eq:def_gLM}, where $P_h^{(0)}(k) = \frac{2 H^2}{M_{\rm pl}^2} k^{-3} = \frac{r}{2} \frac{2\pi^2}{k^3} A_S$ is the isotropic part of the tensor power spectrum parametrised by the tensor-to-scalar ratio $r$ and the amplitude of scalar fluctuations ($A_S = 2.23 \times 10^{-9}$ \cite{Ade:2013zuv}). 
In the present noncommutative scenario, the anisotropic coefficient becomes 
\begin{eqnarray}
g_{LM}
= - 16 \left( \frac{\alpha}{k_0} \right)^2 
\left[ 
\sqrt{\pi} \delta_{L,0} \delta_{M,0}
+ \frac{4\pi}{5}  Y_{LM}^* (\hat{\boldsymbol \alpha}) \delta_{L, 2} \right]
 ~, \label{eq:gLM}
\end{eqnarray}
where $g_{LM}^* = (-1)^M g_{L\, -M}$ holds and we have set $f_0(k) = f_2(k) = (k / k_0)^{-2}$ with $1 / k_0 = \tau_0 = 14 ~ {\rm Gpc}$ being the present horizon scale. In subsection~\ref{subsec:fisher}, we estimate the expected error bar on $g_{LM}$ instead of ${\boldsymbol \alpha}$. Then, note that $g_{00}$ is sensitive to the size of noncommutativity $\alpha = |{\boldsymbol \alpha}|$ alone, while the direction of noncommutativity $\hat{\boldsymbol \alpha}$ is only determined by the $g_{2M}$ measurement.

We notice that this form is appropriate under the perturbative treatment of the anisotropic part and the positiveness of the power spectrum for any direction of ${\boldsymbol \alpha}$, i.e., $\alpha \ll k / 5$. With the projection $\ell \sim k \tau_0$, the corresponding CMB scale is roughly evaluated as $\ell \gg 5 \alpha / k_0$. This means that the CMB computation based on eq.~\eqref{eq:GW_power_n0} is valid for $\alpha / k_0 \ll 2/5$, since the largest CMB scale is $\ell = 2$. This condition is also compatible with a bound evaluated from a constraint on today's graviton mass, namely $\alpha \lesssim 0.01 ~{\rm Mpc}^{-1}$ \cite{Cai:2007xr}. 

\subsection{CMB power spectrum}

The tensor-mode CMB anisotropy is expressed as \cite{Shiraishi:2010sm, Shiraishi:2010kd}
\begin{eqnarray}
a_{\ell m}^X = 
4\pi (-i)^{\ell} \int \frac{d^3 {\bf k}}{(2\pi)^{3}}
 {\cal T}_{\ell}^X(k) \sum_{\lambda = \pm 2} 
\left(\frac{\lambda}{2}\right)^{x} \bar{h}_{\bf k}^{(\lambda)} 
{}_{-\lambda}Y_{\ell m}^*(\hat{\bf k})~, \label{eq:alm}
\end{eqnarray}
where $X = T$ (temperature), $E$ (E-mode polarisation) and $B$ (B-mode polarisation), $\lambda = \pm 2$ denotes the spin of the tensor mode and $x$ represents the parity of the field, namely $x = 0$ for $X = T, E$ and $x = 1$ for $X = B$. ${\cal T}_{\ell}^X(k)$ is the radiation transfer function and expresses the enhancements of the temperature mode for $\ell \lesssim 100$ due to the integrated Sachs-Wolfe effect and two peaks at $\ell \sim 10$ and $100$ in the polarisation modes due to Thomson scattering. We assume that anisotropic effects of the noncommutativity on the transfer function are negligible. Taking the ensemble average of the square of $a_{\ell m}^{X}$ and dealing with the addition of angular momentum as shown in ref.~\cite{Shiraishi:2010kd}, the CMB power spectrum is formulated as 
\begin{eqnarray}
\Braket{a_{\ell_1 m_1}^{X_1} 
a_{\ell_2 m_2}^{X_2 *}}
= C_{\ell_1 \ell_2}^{X_1 X_2}(\gamma = 0)  \delta_{x_1, x_2} \delta_{\ell_1, \ell_2} \delta_{m_1, m_2}
+
\sum_{L = 0,2} C_{\ell_1 m_1 \ell_2 m_2}^{X_1 X_2}(L) ~,
\end{eqnarray}
where the anisotropic power spectrum is 
\begin{eqnarray}
C_{\ell_1 m_1 \ell_2 m_2}^{X_1 X_2}(L) 
&=& i^{\ell_2 -\ell_1}
C_{\ell_1 \ell_2}^{X_1 X_2}(\gamma = -2) 
\delta^{\rm even}_{x_1 + x_2 + \ell_1 + \ell_2}
I_{\ell_1 \ell_2 L}^{-2 2 0}
\nonumber \\ 
&&\times \sum_{M}  
g_{LM}
 (-1)^{m_1} 
 \left(
  \begin{array}{ccc}
  \ell_1 & \ell_2 & L \\
  -m_1 & m_2 & M
  \end{array}
 \right) ~,
\end{eqnarray}
with
\begin{eqnarray}
C_{\ell_1 \ell_2}^{X_1 X_2}(\gamma) 
&\equiv& \frac{2}{\pi}
\int_0^\infty k^2 dk {\cal T}_{\ell_1}^{X_1}(k) {\cal T}_{\ell_2}^{X_2}(k) 
P_h^{(0)}(k) \left( \frac{k}{k_0} \right)^\gamma ~, \\
I^{s_1 s_2 s_3}_{l_1 l_2 l_3} 
&\equiv& \sqrt{\frac{(2 l_1 + 1)(2 l_2 + 1)(2 l_3 + 1)}{4 \pi}}
\left(
  \begin{array}{ccc}
  l_1 & l_2 & l_3 \\
   s_1 & s_2 & s_3 
  \end{array}
 \right) ~, \\
\delta^{{\rm even}}_l 
&\equiv& \begin{cases}
1 & (l = {\rm even}) \\
0 & (l = {\rm odd})
\end{cases} ~.
\end{eqnarray}
The anisotropic power spectrum satisfies $C_{\ell_1 m_1 \ell_2 m_2}^{X_1 X_2 *}(L) = (-1)^{m_1 + m_2} C_{\ell_1 -m_1 \ell_2 -m_2}^{X_1 X_2}(L)$ and $C_{\ell_1 m_1 \ell_2 m_2}^{X_1 X_2}(L) = C_{\ell_2 m_2 \ell_1 m_1}^{X_2 X_1 *}(L)$. The selection rules in $\delta^{\rm even}_{x_1 + x_2 + \ell_1 + \ell_2}$ and $I_{\ell_1 \ell_2 L}^{-2 2 0}$ allow nonzero $TT$, $TE$, $EE$ and $BB$ in $\ell_2 = \ell_1$ and $\ell_1 \pm 2$, and $TB$ and $EB$ in $\ell_2 = \ell_1 \pm 1$. We can also notice that $TB$ and $EB$ vanish when $L=0$ because rotational invariance, namely $\delta_{\ell_1, \ell_2} \delta_{m_1, m_2}$, is kept. These are common signatures of the primordial quadrupole anisotropy in the CMB power spectra \cite{Watanabe:2010bu, Soda:2012zm}. 

\begin{figure}
  \begin{center}
    \includegraphics[width = 12 cm]{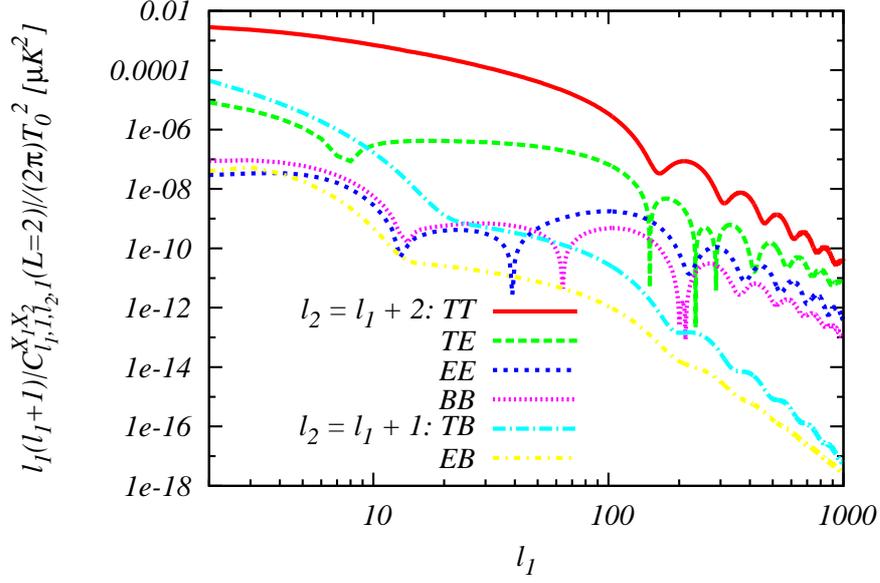}
  \end{center}
  \caption{CMB quadrupole power spectrum: $C_{\ell_1, 1, \ell_2, 1}^{X_1 X_2}(L = 2)$ for $\ell_2 = \ell_1 + 2$ ($TT$, $TE$, $EE$ and $BB$) and $\ell_2 = \ell_1 + 1$ ($TB$ and $EB$). Here, we set $r = 10^{-2}$ and $\alpha / k_0 = 0.1$, and fix ${\boldsymbol \alpha}$ to be along $z$ axis. Then, $g_{2M} \approx -0.25 \delta_{M,0}$ holds.}
  \label{fig:Cl}
\end{figure} 

Figure~\ref{fig:Cl} depicts auto- and cross-correlated power spectra of the temperature and polarisations generated from the quadrupole anisotropy: $C_{\ell_1 m_1 \ell_2 m_2}^{X_1 X_2}(L = 2)$ for $m_1 = m_2 = 1$. Here, the direction of ${\boldsymbol \alpha}$ is fixed to be parallel to the $z$ axis and hence $Y_{LM}(\hat{\boldsymbol \alpha}) = \sqrt{\frac{2L+1}{4\pi}} \delta_{M,0}$ holds. In this figure, it is obvious that compared with the usual scale-invariant case, the CMB power spectra have highly red-tilted shapes due to the additional $k^{-2}$ dependence of the quadrupole anisotropy. One may also notice that the two parity-odd correlations ($TB$ and $EB$) decay more rapidly than the other four parity-even ones. Due to this scale dependence, the expected error on $g_{LM}$ converges at very low $\ell$ as shown in the next subsection.

\subsection{Error estimation}\label{subsec:fisher}

As in the present noncommutative case, when the statistical anisotropy in the CMB power spectrum is regarded as a tiny modulation of the scale-invariant isotropic component, it can be well-estimated by using the QML estimator \cite{Hanson:2009gu, Hanson:2010gu}. Then, the Fisher matrix for $g_{LM}$ can be written as
\begin{eqnarray}
{\cal F}_{LM, L' M'} 
= 
\sum_{i,j}
\frac{\delta {\bf C}^{i}(L)}{\delta g_{LM}^*} 
  ({\bf Cov}^{-1})^{ij}
\left( \frac{\delta {\bf C}^{j}(L')}{\delta g_{L'M'}^*} \right)^* ~,
\end{eqnarray}
where $i$ and $j$ runs over given sets of auto- and cross-correlated power spectra, ${\bf C}^{i}$ is the $i$th CMB power spectrum and ${\bf Cov}^{ij}$ is the covariance matrix element given by the $i$th and $j$th power spectra. For simplicity, let us ignore off-diagonal elements in the observed power spectra, i.e., $\Braket{a_{\ell_1 m_1}^{\rm obs} a_{\ell_2 m_2}^{{\rm obs} *}} \approx \tilde{C}_{\ell_1} \delta_{\ell_1, \ell_2} \delta_{m_1, m_2}$. Furthermore, on the basis of the fact that $C_{\ell_1 \ell_2}^{X_1 X_2}(\gamma) \approx C_{\ell_1 \ell_2}^{X_2 X_1}(\gamma)$ for $\ell_2 - 2 \leq \ell_1 \leq \ell_2+2$, we can obtain that $C_{\ell_1 m_1 \ell_2 m_2}^{X_1 X_2}(L) \approx (-1)^{m_1+m_2} C_{\ell_2 -m_2 \ell_1 -m_1}^{X_1 X_2}(L)$. After these two approximations the Fisher matrix is reduced to 
\begin{eqnarray} 
{\cal F}_{LM, L' M'} 
&\approx& \sum_{\ell_1 m_1 \ell_2 m_2} 
\sum_{\substack{i \leftrightarrow X_1 X_2 \\ j \leftrightarrow X_1' X_2'}}
\frac{\delta C_{\ell_1 m_1 \ell_2 m_2}^{i}(L)}{\delta g_{LM}^*} 
  (Cov^{-1})_{\ell_1 \ell_2}^{ij}
\left( \frac{\delta C_{\ell_1 m_1 \ell_2 m_2}^{j}(L')}{\delta g_{L'M'}^*} \right)^* \nonumber \\ 
&=& 
\sum_{\ell_1 \ell_2} 
\sum_{\substack{i \leftrightarrow X_1 X_2 \\ j \leftrightarrow X_1' X_2'}}
\delta^{\rm even}_{x_1' + x_2' + \ell_1 + \ell_2}
\delta^{\rm even}_{x_1 + x_2 + \ell_1 + \ell_2}
(I_{\ell_1 \ell_2 L}^{-2 2 0} )^2
\frac{\delta_{L L'} \delta_{M M'}}{2L+1} \nonumber \\ 
&&\times C_{\ell_1 \ell_2}^{i}(\gamma = -2) 
(Cov^{-1})_{\ell_1 \ell_2}^{ij} 
C_{\ell_1 \ell_2}^{j}(\gamma = -2) ~,
\end{eqnarray}
with 
\begin{eqnarray}
Cov_{\ell_1 \ell_2}^{ij} 
&=& 
\tilde{C}_{\{ \ell_1}^{X_1 X_1'} \tilde{C}_{\ell_2\} }^{X_2 X_2'} 
+ \tilde{C}_{\{ \ell_1}^{X_1 X_2'} \tilde{C}_{\ell_2\} }^{X_2 X_1'}
  ~,
\end{eqnarray}
where $A_{\{\ell_1} B_{\ell_2\}} \equiv \frac{1}{2}[A_{\ell_1} B_{\ell_2} + A_{\ell_2} B_{\ell_1} ]$ denotes the symmetrization under the permutation of $\ell_1$ and $\ell_2$. 

From now, we consider error estimations from three auto-correlations ($TT$, $EE$ and $BB$), a combination of the temperature and E-mode anisotropies ($TT+TE+EE$) and a parity-odd combination ($TB+EB$). The covariance matrix involves the observed power spectrum $\tilde{C}_{\ell}$ that is the sum of the signal and the instrumental noise. In this paper, let us consider the noise information of the {\it Planck} and the proposed PRISM experiments \cite{Ade:2013ktc, Andre:2013afa, Shiraishi:2013vha}. As the temperature and E-mode signals, we adopt the CMB power spectra consistent with the {\it Planck} data~\cite{Ade:2013zuv}. On the other hand, the observed B-mode spectrum is still unknown and here we simply assume that the B-mode signal is consistent with the scale-invariant isotropic power spectrum $C_{\ell \ell}^{BB}(\gamma = 0)$. Recall that this magnitude is proportional to the tensor-to-scalar ratio $r$. Moreover, for simplicity, we adopt a null hypothesis for the observed TB and EB correlations, i.e., $\tilde{C}_{\ell}^{TB} = \tilde{C}_{\ell}^{EB} = 0$. 

\begin{figure}[t]
  \begin{tabular}{cc}
    \begin{minipage}{0.5\hsize}
  \begin{center}
    \includegraphics[width=1\textwidth]{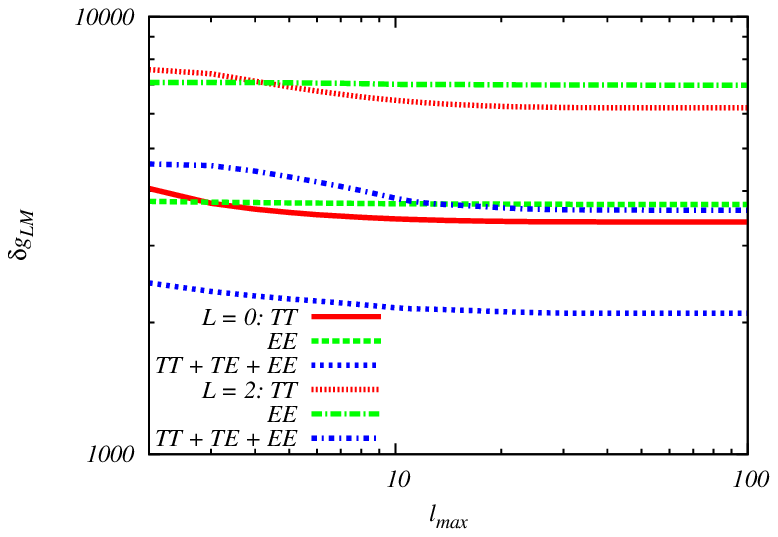}
  \end{center}
\end{minipage}
\begin{minipage}{0.5\hsize}
  \begin{center}
    \includegraphics[width=1\textwidth]{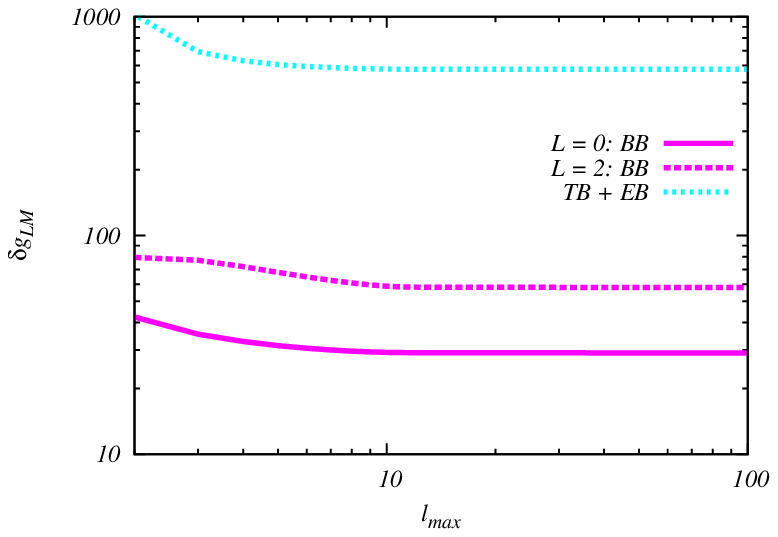}
  \end{center}
\end{minipage}
\end{tabular}
  \caption{$\delta g_{00}$ and $\delta g_{2M}$ estimated from $TT$, $EE$, $TT + TE + EE$ (left panel), $BB$ and $TB + EB$ (right panel) in a cosmic-variance-limited experiment. Here, we take $r = 10^{-2}$.} \label{fig:error}
\end{figure} 

Figure~\ref{fig:error} describes the numerical results of $1\sigma$ errors on $g_{LM}$ given by 
\begin{eqnarray}
\delta g_{LM} = {\cal F}_{LM, LM}^{-1/2}~.
\end{eqnarray} 
$\delta g_{LM}$ does not depend on $M$ because in the analysis we ignore the off-diagonal elements in the covariance matrix. In figure~\ref{fig:error}, we assume a cosmic-variance-limited experiment. One can observe from this figure that $\delta g_{LM}$ converges before $\ell_{\rm max} \sim 100$ owing to the highly red-tilted CMB spectra as shown in figure~\ref{fig:Cl}. It is obvious that when $r = 10^{-2}$, $\delta g_{LM}$ exceeds $10^3$ in the analysis with the temperature or E-mode polarisation, while the analysis with the B-mode polarisation reduces $\delta g_{LM}$ by one or two orders of magnitude. This is a consequence that the cosmic variances in the temperature and E-mode polarisation are much larger than the B-mode cosmic variance with $r = 10^{-2}$. In the measurements of the temperature and E-mode polarisation, because the theoretical power spectrum $C_{\ell_1 \ell_2}^{X_1 X_2}(\gamma)$ is proportional to $r$ and there is no $r$ dependence in the covariance matrix, a simple scaling relation $\delta g_{LM}^{TT} \propto \delta g_{LM}^{EE} \propto \delta g_{LM}^{TT+TE+EE} \propto r^{-1}$ holds. Meanwhile, if involving the B-mode information, the covariance matrix also depends on $r$. Especially, in a cosmic-variance-limited experiment, the covariance matrix is determined by $r$ alone, and hence we obtain $\delta g_{LM}^{BB} \propto r^0$ and $\delta g_{LM}^{TB+EB} \propto r^{-1/2}$ (see also table~\ref{tab:error_B}). The results in figure~\ref{fig:error} and these magnitude relations lead to the conclusion that, for $r < 10^{-2}$, namely $r$ allowed by current observations, the most stringent constraint on $g_{LM}$ comes from $BB$. We can also notice a rough relation $\delta g_{2M} \sim 2 \delta g_{00}$ in all cases. 

\begin{table}[t]
\begin{center}
  \begin{tabular}{|c||c|c|c|c|c|c|c|c|} \hline
    & \multicolumn{2}{|c|}{$BB$} & \multicolumn{2}{|c|}{$TB+EB$}  \\ \hline
    $r$ & $10^{-2}$ & $10^{-4}$ & $10^{-2}$ & $10^{-4}$  \\ \hline
    {\it Planck} & 180 (87) & 9100 (4600) & 1200 & 92000 \\ 
    PRISM & 73 (36) & 160 (78) & 700 & 11000  \\
    ideal & 58 (30) & 58 (30) & 580 & 5800 \\ \hline
  \end{tabular}
\end{center}
\caption{$\delta g_{2M}$ (and $\delta g_{00}$) estimated from $BB$ and $TB+EB$ assuming $r = 10^{-2}$ and $10^{-4}$. Here, we take into account the effect of $70\%$ sky coverage ($f_{\rm sky} = 0.7$) in the {\it Planck} and PRISM cases by following $\delta g_{LM} \propto f_{\rm sky}^{-1/2}$. Note that we only describe $\delta g_{2M}$ in the $TB + EB$ analysis because of the absence of $L=0$ signals. 
In the ideal experiment, $\delta g_{LM}$ from $TB + EB$ is exactly proportional to $r^{-1/2}$, while for $BB$ it is independent of $r$.}\label{tab:error_B}
\end{table}

Table~\ref{tab:error_B} presents $\delta g_{LM}$ estimated from the B-mode polarisation ($BB$ and $TB + EB$) under the {\it Planck}, PRISM and ideal cosmic-variance-limited experiments. We take $r = 10^{-2}$ and $10^{-4}$ for comparison. This table shows that the {\it Planck} measurement cannot detect $g_{LM}$ less than ${\cal O}(10-100)$ except in the $BB$ analysis with $r = 10^{-2}$. In contrast, the $BB$ analysis by the PRISM experiment will measure it at 68 \% CL even if $r$ is very small (e.g., $r = 10^{-4}$). In case that $r$ is large, $TB + EB$ may also become a good observable. 

The range of $\alpha$ coming from our perturbative treatment \eqref{eq:GW_power_n0}, namely $\alpha / k_0 \ll 2/5$, corresponds to $g_{LM}$ satisfying $|g_{LM}| \lesssim {\cal O}(1)$. The above results indicate that such small $\alpha$ cannot be measured.

\section{Conclusion}

Noncommutativity is a key indicator of non-standard high energy physics and it may imprint interesting effects on the primordial fluctuations. This paper has focused on signatures of primordial gravitational waves affected by the noncommutativity of fields in the observed CMB fluctuations. 

In addition to the usual isotropic (nearly) scale-invariant power spectrum of gravitational waves, the noncommutativity gives two contributions with $k^{-5}$ dependence: an isotropic and a quadrupolar ones. Hence, the resulting CMB power spectra are highly red-tilted and have non-vanishing off-diagonal components, i.e., $\ell_2 = \ell_1 \pm 2$ in $TT$, $TE$, $EE$ and $BB$, and $\ell_2 = \ell_1 \pm 1$ in $TB$ and $EB$. 

By applying the QML estimator, we have estimated the expected uncertainty on the statistical anisotropy. We have used the usual spherical harmonic parametrisation with coefficients $g_{LM}$ and computed the $1\sigma$ error bars $\delta g_{LM}$ through the analyses of auto-correlations, i.e., $TT$, $EE$ and $BB$, and combined analyses with auto- and cross-correlations, i.e., $TT + TE + EE$ and $TB + EB$. Then, we have found that $\delta g_{LM}$ converges before $\ell \sim 100$ because of the highly red-tilted spectrum shapes. We have confirmed that, since the B-mode cosmic variance is much smaller than the temperature and E-mode ones on such large scales, $BB$ is most informative to measure $g_{LM}$. Potentially, we can measure $g_{00} = 30$ and $g_{2M} = 58$, where $M = 0, \pm 1, \pm 2$, at 68\% CL in a cosmic-variance-limited experiment. It is hard to measure $g_{LM}$ with {\it Planck} due to the lack of sensitivity to polarisations, while a PRISM like experiment may attain such accuracy level. 

Our analysis has showed that if the noncommutative parameter $\alpha$ is enough small for the perturbative treatment to be justified, it is impossible to detect it even in the ideal noise-free experiment. On the other hand, there may exist theoretical models predicting detectable tensor $g_{LM}$. The formalism and methodology for dealing with the CMB statistical anisotropy developed in the present paper will be applicable to future phenomenological studies on such models.

\acknowledgments
We thank Michele Liguori for useful comments. MS was supported in part by a Grant-in-Aid for JSPS Research under Grant No.~25-573. DFM thanks Research Council of Norway. This work was partly supported in part by the ASI/INAF Agreement I/072/09/0 for the Planck LFI Activity of Phase E2.

\appendix

\bibliography{paper}
\end{document}